\documentclass[%
12pt,
]{iopart}

\usepackage{graphicx}
\usepackage{iopams}
\usepackage{siunitx}
\usepackage{booktabs}
\usepackage{multirow}

\renewcommand*{\rmd}{{\rm d}}
\renewcommand*{\rme}{\mathrm{e}}
\renewcommand*{\mr}[1]{\mathrm{#1}}
\renewcommand*{\bs}[1]{\boldsymbol{#1}}

\newcommand*{\kb}{k_\mathrm{B}}

\newcommand*{\Te}{T_\mathrm{e}}

\begin{document}

%-----TITLE
\title{Molecular dynamics study of electronic temperature effects on the laser ablation of silicon}

%-----AUTHORS
\author{Ryo Kobayashi$^1$ and Tomohito Otobe$^2$}
\address{$^1$ Department of Physical Science and Engineering, Nagoya Institute of Technology, Gokiso-cho, Showa-ku, Nagoya 466-8555, Japan}
\address{$^2$ Kansai Photon Science Institute, National Institutes for Quantum Science and Technology (QST), Kyoto 619-0215, Japan}
\ead{kobayashi.ryo@nitech.ac.jp}

%-----ABSTRACT
\begin{abstract}
 The molecular dynamics (MD) approach is an effective tool for investigating atomistic dynamical phenomena at the surface of materials under strong laser irradiation.
 Therefore, numerous laser ablation MD simulation studies have been conducted to date.
 However, in most MD studies, non-thermal and entropic effects via hot electrons on interatomic interactions that could cause significant differences in the simulation results are not considered.
 In this study, the MD simulation of the laser ablation of the Si surface was conducted using an interatomic potential whose parameters depended on the electronic temperature. 
 Moreover, the results obtained with and without electronic temperature dependence were compared.
 The electronic temperature dependence resulted in an approximately four-times-greater compressive pressure near the surface, enhanced evaporation of atomic or smaller clusters, and slightly longer melt depth.
 Compared to the strong compressive pressure near the surface, the tensile pressure, which originated from the reflection of the compressive pressure wave at the surface, and ablation phenomena were less dependent on the electronic temperature.
\end{abstract}

%-----PACS number
% 07.05.Tp: Computer modeling and simulation
% 34.20.Cf: interatomic potentials
% 31.15.B-: approximate calculation for ab initio calculations
% 61.20.Ja: MD calculation in liquid structure modeling
% \pacs{07.05.Tp, 34.20.Cf, 31.15.B-, 61.20.Ja}
% \keywords{laser ablation; electronic temperature; two-temperature model; molecular dynamics}

\maketitle

 \section{Introduction}\label{sec:intro}

 The response of materials to strong laser pulses has been attracting research interest because it is the foundation of material processing using lasers such as
 surface nanostructure manufacturing,\cite{Bonse2002-kw,Rudenko2017-hu} laser shock peening,\cite{Fabbro1990-xr} and pulsed laser deposition.\cite{M_Castillerjo_P_M_Ossi_L_Zhigilei_undated-uw}
 To precisely understand these phenomena, extensive experimental research has been conducted to observe the response of materials with respect to several laser parameters, such as the wavelength, pulse duration, and fluence.\cite{Dieleman1992-rn,M_Castillerjo_P_M_Ossi_L_Zhigilei_undated-uw,Stafe_undated-dv}
 However, because the phenomena are intrinsically multi-scale in length and time and multi-physics, it is difficult to fully understand the elemental physics influencing the phenomena.
 Therefore, there is no standard physical model that can predict material response (e.g., ablation yield, ablation depth, damage layer depth, and residual stress) consisting of the parameters of the incident laser and material.
 Particularly, the current understanding of atomistic-scale, fast, and non-equilibrium processes occurring at the material surface is far from sufficient, since these processes are difficult to directly observe experimentally.
 
 Computational approaches have been employed to study the elemental processes of laser-material interactions,\cite{Miloshevsky2022-or} including the Vlasov equation\cite{Tani2021-kk} or time-dependent density functional theory (TD-DFT) calculations\cite{Kumada2016-jn} for laser absorption by electrons in materials, electromagnetic-hydrodynamic simulation for the meso-to-macroscopic morphological evolution of the surface.\cite{Rudenko2017-hu,Rudenko2019-dn}
 The molecular dynamics (MD) approach is useful for studying dynamic non-equilibrium phenomena of materials under strong perturbations from an atomistic viewpoint and is thus successfully applied in laser ablation studies.\cite{Perez2003-qx,Lorazo2006-ga,M_Castillerjo_P_M_Ossi_L_Zhigilei_undated-uw}
 To simulate the interaction between the laser and materials in MD, a combination of the two-temperature model (TTM) and MD is widely used.\cite{Zhigilei2009-cv,Ivanov2003-ae}
 The TTM solves two thermal diffusion equations for electronic and lattice temperatures by considering the energy transfer between the two systems.
 TTM-MD replaces the lattice part of the diffusion equation with MD and enables the treatment of drastic changes in atomic structures under strong laser irradiation.
 Thus, in this study, the TTM-MD approach was used for the atomistic simulation of laser ablation.

 In TTM-MD, the electronic temperature, $\Te$, increases by absorbing the energy of the incident laser.
 The energy is then transferred and converted to the lattice temperature as a result of the difference between electronic and lattice temperatures, and no other effect originating from hot electrons is included.
 Consequently, non-thermal contributions that are significant especially in the case of short-pulse lasers are not considered.\cite{Rousse2001-xt}
 Since the non-thermal and entropic effects of hot electrons on interatomic potentials (IPs) are considered to be significant near the surface irradiated by the strong laser,\cite{Tanaka2018-ln,Lee2019-pk} there are several studies on the contribution of changes in the IPs by hot electrons to the laser-material interaction.\cite{Tanaka2022-ec1,Lee2019-pk,Zhang2020-mj,Shokeen2010-nu,Shokeen2013-gf,Darkins2018-ny,Bauerhenne2019-cd}
 However, the influence of the $\Te$ dependence of IP on the laser ablation phenomena is not entirely clear, because large-scale laser-ablation MD studies with the $\Te$-dependent IPs and spatiotemporally-changing $\Te$ have not been conducted intensively.
 Thus, this study aimed to investigate the effects of hot electrons induced by laser irradiation on the ablation phenomena at the material surface.

 The IP directly influences the reliability of MD simulation results.
 Thus far, most IPs used for large-scale laser-ablation MD studies have been constructed to reproduce the properties of materials at the ground state or under the equilibrium condition.
 There have been several reports on the construction of IPs considering the $\Te$ effect, such as IPs for metals\cite{Tanaka2022-ec1,Zhang2020-mj} and Si.\cite{Shokeen2010-nu,Darkins2018-ny,Bauerhenne2019-cd}
 This study employed the IP for Si proposed by Shokeen and Shelling\cite{Shokeen2010-nu} that depends on the temperature of electrons associated with Si atoms.
 TTM-MD provides the $\Te$ at a certain Si atom by interpolating those on mesh grids.
 By comparing the results of the TTM-MD simulation with and without the $\Te$ dependency on the IP, we extracted the contributions of $\Te$ on the laser ablation phenomena of the Si surface.
 Section \ref{sec:method} describes TTM-MD and $\Te$-dependent IP for Si.
 The simulation results and $\Te$ dependence are discussed in Section~\ref{sec:results}.
 Finally, Section~\ref{sec:conclusion} summarises the results and concludes the study.

 \section{Method}\label{sec:method}

  \subsection{TTM-MD}\label{sec:TTM-MD}

  \begin{figure*}[tbp]
   \centering
   \includegraphics[keepaspectratio,width=\textwidth]{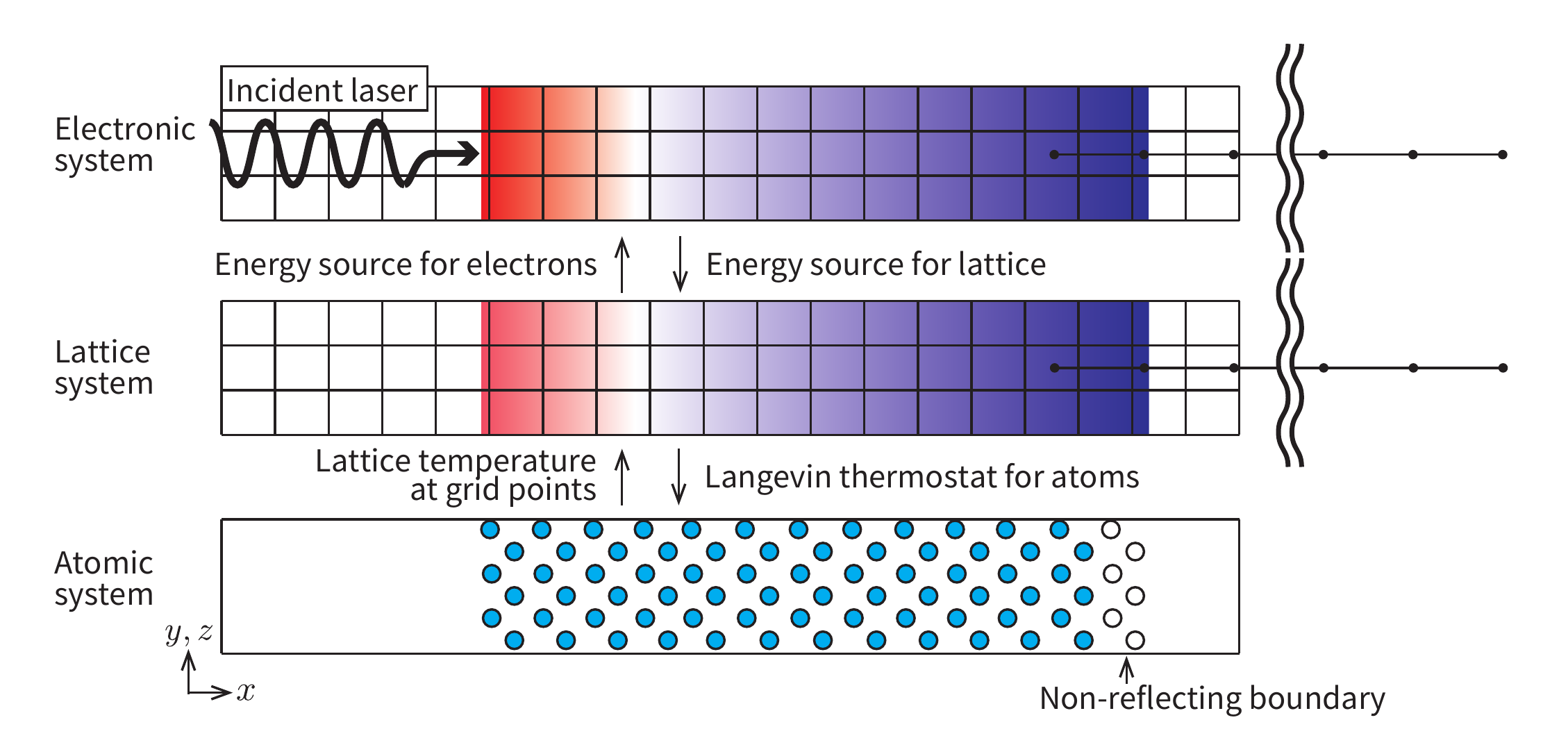}
   \caption{Schematic of TTM-MD simulation. Temperature evolution in electronic and lattice systems is solved with an orthogonal grid, and atom motions are treated by MD. 
   The energy transfer from the atom system to the electron system is achieved by TTM, and the opposite is achieved by the Langevin thermostat for the atom system. 
   The non-reflecting boundary condition is set to a few atomic layers of the other side of the incident laser.}
   \label{fig:sim-config}
  \end{figure*}

  The TTM-MD method\cite{Ivanov2003-ae,Zhigilei2009-cv} was employed for MD simulation under strong laser irradiation.
  The 3D TTM was used for the region with the atoms, and the 1D TTM treated the heat dissipation at the deeper region where atoms were not clearly observed as shown in Fig.~\ref{fig:sim-config}.
  In the 3D TTM region, $\Te$ is expressed by the three-dimensional thermal diffusion equation using an orthogonal grid, as follows:
  \begin{equation}
   \rho_\mr{e}C_\mr{e}\frac{\partial \Te}{\partial t} 
    = \nabla \left[ \kappa_\mr{e} \nabla \Te\right] - g_\mr{p} (\Te -T_\mr{l}) +g_\mr{s} T'_\mr{l} +S(x,t),
    \label{eq:3d-ttm}
  \end{equation}
  where $\rho_\mr{e}$ is the electron density, $C_\mr{e}$ is the electron heat capacity, $\kappa_\mr{e}$ is the electron thermal conductivity, $g$ is the electron-phonon coupling factor, and $S(x,t)$ is the source term describing the laser irradiation per unit area and per unit time.
  The lattice temperature $T_\mr{l}$ of a grid point is computed as the local average of the kinetic energies of atoms inside a cell containing the grid point.
  Energy transfer between the electronic and lattice systems is achieved by the second term on the right hand side in Eq.~\ref{eq:3d-ttm}.
  The transfer from the lattice system to the atomic system is described by the Langevin thermostat as:
  \begin{equation}
   m_i \ddot{\bs{r}}_i(t) = \bs{F}_i(t) - m_i \gamma_i \tilde{\bs{v}}_i(t) +\bs{R}(t),
    \label{eq:Langevin}
  \end{equation}
  where $m_i$, $\bs{F}_i(t)$, and $\bs{R}(t)$ are the atomic mass of atom-$i$, atomic force on atom-$i$, and fluctuating force, respectively.
  $\tilde{v}_i$ is the thermal velocity obtained by subtracting the average translational motion of corresponding cell $c$ as,
  $\displaystyle \tilde{\bs{v}}_i = \bs{v}_i -\sum_{j\ \text{in}\ c} \bs{v}_j/n_c$, where $n_c$ denotes the number of atoms in the cell $c$.
  The friction coefficient is determined as:
  \begin{equation}
   \gamma_i = 
    \left\{\begin{array}{ll}
               \gamma_\mr{p} +\gamma_\mr{s}, & \text{for}\ |v_i| > v_\mr{th}, \\
               \gamma_\mr{p}, & \text{otherwise,}
           \end{array}
      \right.
  \end{equation}
  where $\gamma_\mr{p}$ and $\gamma_\mr{s}$ are the friction coefficients originating from by the electron-phonon coupling and electron stopping, respectively.
  The electron stopping works only when the velocity $v_i$ exceeds the threshold velocity $v_\mr{th}$.
  $T'_\mr{l}$ is the temperature of the atoms of $v_i > v_\mr{th}$.
  The electron-phonon coupling factor $g_\mr{x}$ is determined using $\gamma_\mr{x}$, the cell volume, and the number of atoms in the cell to balance the input and output of energies between the electronic and atomic systems.\cite{Ivanov2003-ae}
  
  In the 1D TTM region, MD is not performed explicitly, and instead, the lattice thermal diffusion equation is solved.
  Considering $x$ to be the one-dimensional direction, the electronic and lattice thermal diffusion equations are written as:
  \begin{eqnarray}
   \rho_\mr{e}C_\mr{e}\frac{\partial \Te}{\partial t} 
    &=& \frac{\rmd}{\rmd x} \left[ \kappa_\mr{e} \frac{\rmd \Te}{\rmd x}\right] - g_\mr{p} (\Te -T_\mr{l}) +S(x,t), \label{eq:1d-Te} \\
   \rho_\mr{l}C_\mr{l}\frac{\partial T_\mr{l}}{\partial t} 
    &=& \frac{\rmd}{\rmd x} \left[ \kappa_\mr{l} \frac{\rmd T_\mr{l}}{\rmd x}\right] + g_\mr{p} (\Te -T_\mr{l}),\label{eq:1D-Tl}
  \end{eqnarray}
  where $\rho_\mr{l}$, $C_\mr{l}$ and $\kappa_\mr{l}$ are the density, heat capacity, and thermal conductivity of the lattice system, respectively.
  These 1D and 3D TTMs are connected at around the $x$-position of the atoms working as the non-reflecting boundary condition for the atomic system.
  The details of the connection algorithm are provided in the Appendix.

  %...Laser source S(x,t)
  The laser energy absorbed by the Si surface was modelled as the instantaneous increase of electronic temperature whose intensity exponentially decays with depth $x$ according to the Beer-Lambert law, as follows:
  \begin{equation}
   S(x,t) = g(t) I_0 \rme^{-x/l_\mr{skin}},
  \end{equation}
  where $I_0$ and $l_\mr{skin}$ are the intensity immediately inside the surface and the penetration depth, respectively.
  An incident laser pulse was assumed to have a Gaussian temporal shape $g(t)$, and the pulse duration $\tau$ was defined as the full width at half maximum of the Gaussian.
  This study only considered the pulse duration $\tau$ of 100 fs.

  %...Non-reflecting boundary condition
  The present MD system employed a non-reflecting boundary condition (NRBC) suggested by Shugaev et al.\cite{Shugaev2017-as} at the opposite side of the laser injection,
  as shown in Fig.~\ref{fig:sim-config}.
  In the NRBC, a control parameter, the length of the NRBC region $L^\mr{NR}$, affects the reflectance of the elastic wave at the boundary.
  As the region becomes larger, the damping coefficient in the NRBC decreases and the reflectance consequently decreases.\cite{Kobayashi2011-bi}
  In this study, $L^\mr{NR}$ was set to 3 nm, which corresponds to the width of six unit cells and is sufficient for suppressing the artificial reflection, as detailed in Section~\ref{sec:results}.

  \subsection{$\Te$-dependent Tersoff potential}\label{sec:Te-dep-tersoff}

  %...Te-dep. Tersoff by Shokeen et al. (1995)
  This study employed the $\Te$-dependent Tersoff potential developed by Shokeen and Schelling.\cite{Shokeen2010-nu}
  This potential is based on the modified Tersoff potential formalism by Kumagai {\itshape et al.},\cite{Kumagai2007-if} and 
  its parameters are dependent on the electronic temperature.
  The parameters were optimised to reproduce the cohesive energies, lattice constants, and bulk moduli of Si obtained by the finite-temperature density functional theory (FT-DFT) calculations at several electronic temperatures up to $\kb \Te = 2.5$ eV. 
  The potential was implemented in an open-source program, nap,\cite{Kobayashi2021-gj}
  and confirmed the $\Te$ dependence of the lattice constant of bulk Si.

\begin{figure}[tbp]
   \centering
   \includegraphics[keepaspectratio,width=.4\textwidth]{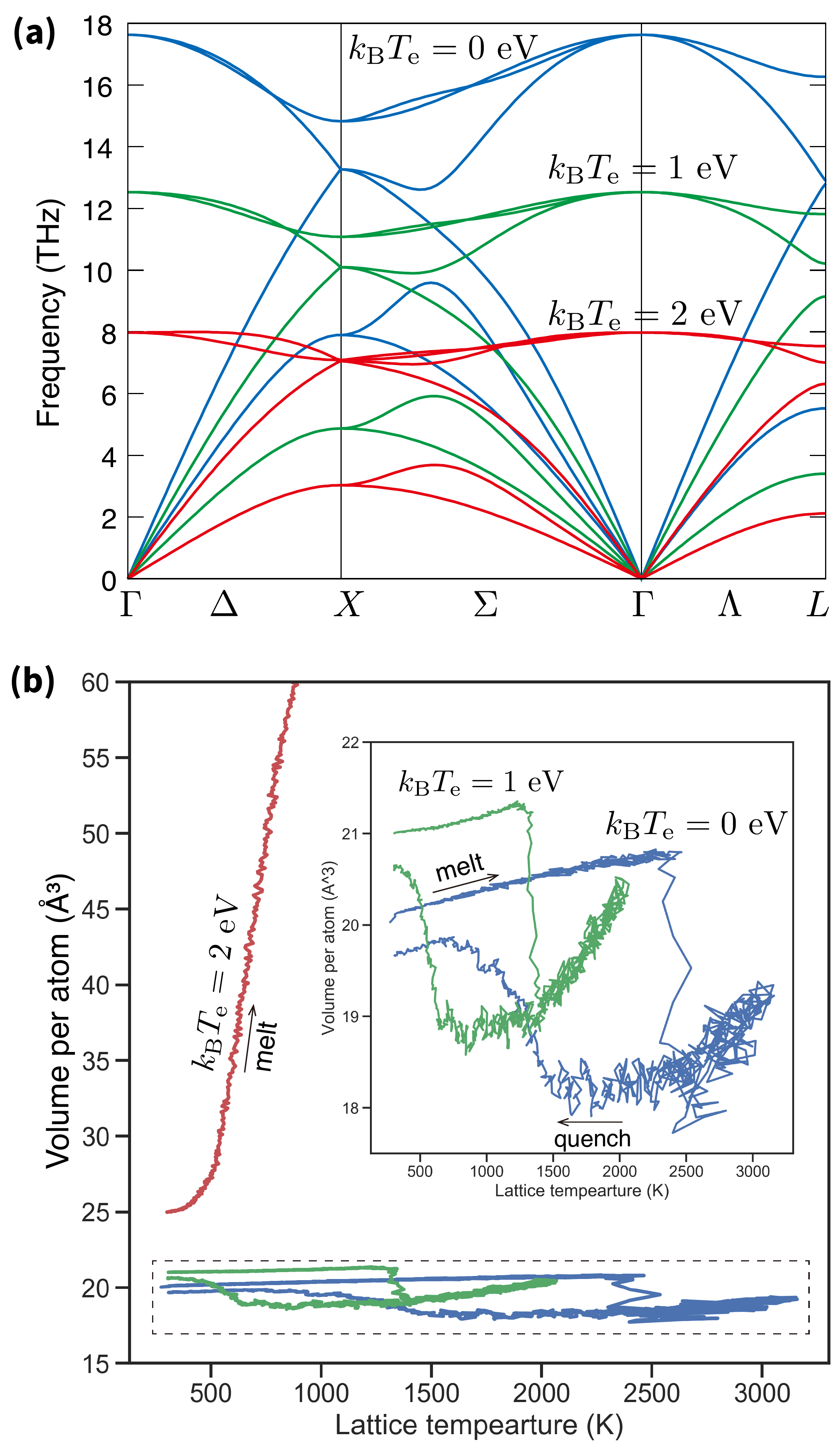}
   \caption{(a) Phonon dispersion relationships and 
   (b) volume (per atom) vs. lattice temperature curves obtained by melt-quench MD using the $\Te$-dependent Tersoff potential. 
   Both plots compare the cases of the electronic temperatures $\kb \Te= 0,\ 1,\ \text{and}\ 2$ eV.}
   \label{fig:Tersoff}
  \end{figure}

  In addition to the static properties shown in the original paper,\cite{Shokeen2010-nu}
  some dynamical properties of the $\Te$-dependent potential were examined.
  Figure \ref{fig:Tersoff}(a) shows the phonon dispersion relationships obtained at different electronic temperatures ($\kb \Te =$ 0, 1, and 2 eV), indicating that the interatomic interaction weakens with increasing $\Te$. 
  Even at $\kb \Te=2$ eV, the diamond structure is stable at 0 K, as there is no imaginary phonon branch.
  To investigate the $\Te$ effects of dynamical behaviour at higher lattice temperatures, 
  we performed melt-quench $NpT$-MD by controlling the temperature $T_\mr{a}$ between 300 and 3,000 K under different $\Te$ conditions.
  Figure \ref{fig:Tersoff}(b) shows the volume of the MD system as a function of temperature $T_\mr{a}$, indicating that the melting temperature decreases with increasing $\Te$.
  Further, at the electronic temperature $\Te=2$ eV, the system evaporates at $T_\mr{a} > 500$ K.
  This strong $\Te$ dependency of the system is considered to affect the response to the strong and short-pulse laser irradiation.

  \subsection{Simulation setup}\label{sec:sim-setup}

  %...System configuration
  The system was quasi-one dimensional in $x$-direction over 800 nm (except for the vacuum region) comprising an approximately 200-nm-long 3D TTM-MD system and
  an approximately 600-nm-long 1D TTM system.
  The 3D atomic system was composed of $350 \times 10 \times 10$ cubic diamond unit cells, and the Miller index of the surface facing the incident laser was (100).
  The periodic boundary condition was used along the $y$ and $z$-directions, whereas the free boundary condition was applied to the $x$-direction.
  The temperature of the right-most grid point in the 1D TTM system was maintained at 300 K.
  The size of the 3D-TTM grid was set to be the same as that of the $2\times 2\times 2$ unit cells, and the 1D-TTM grid size was 2 nm.
  The parameters for the electronic system of Si were acquired from the literature and are listed in Table~\ref{tab:params}.
  Additionally, the $\Te$ dependency of $C_\mr{e}$ proposed by Jay {\itshape et al.}\cite{Jay2017-oz} was used.
  
  \begin{table}[tb]
   \centering
   \caption{Parameters used in the 1D and 3D TTM simulation.}
   \begin{tabular*}{.45\textwidth}{@{\extracolsep{\fill} }lrl}
    \toprule
    Parameter         & Value  & Unit \\
    \midrule
    $\rho_\mr{e}$   & 0.05  &  $\mr{1/\si{\angstrom}^3}$  \\ 
    $D_\mr{e}$ & 20,000  & $\mr{\si{\angstrom}^2/ps}$ \\
    $\gamma_\mr{p}$  & 1.741  &  g/(mol$\cdot$ps) \\
    $\gamma_\mr{s}$  & 39.23  &   g/(mol$\cdot$ps)\\
    $v_\mr{th}$   & 76.76  &  \si{\angstrom}/fs  \\
    $C_\mr{l}$  &  2.06 $\times 10^{-4}$  &  eV/$\mr{K}$ \\
    $\rho_\mr{l}$   & 0.05  &  1/$\mr{\si{\angstrom}^3}$ \\
    $D_\mr{l}$   &  8.8    &   \si{\angstrom}$^2$/ps \\
    \bottomrule
   \end{tabular*}
   \label{tab:params}
  \end{table}

 \section{Results and discussion}\label{sec:results}

  \subsection{Laser ablation by TTM-MD}\label{sec:laser-ttm}

  \begin{figure}[tbp]
   \centering
   \includegraphics[keepaspectratio,width=.5\textwidth]{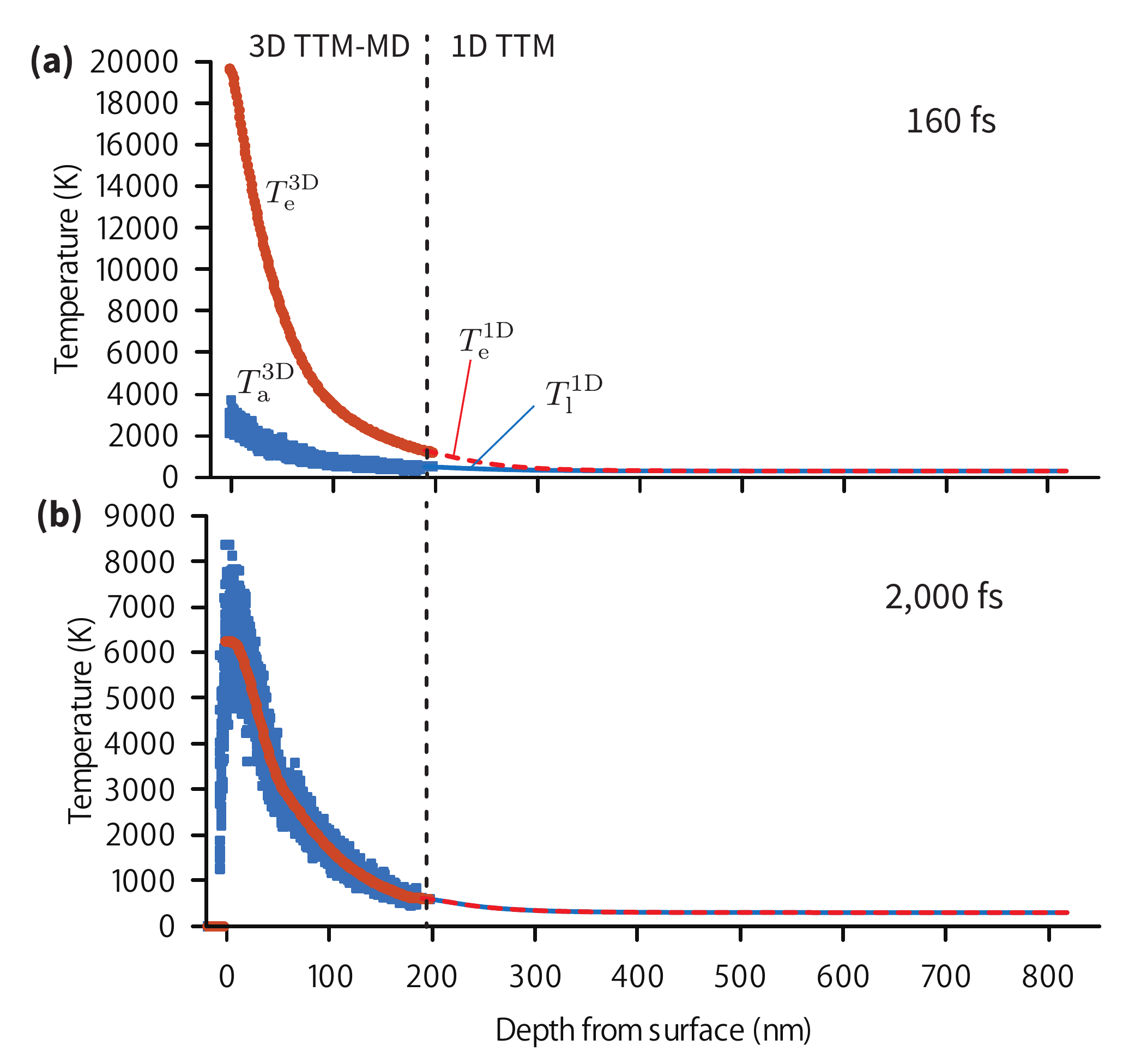}
   \caption{Spatial distribution of temperatures $\Te^\mr{3D}$, $T_\mr{a}^\mr{3D}$, $\Te^\mr{1D}$ and $T_\mr{l}^\mr{1D}$ along $x$-direction at (a) 160 and (b) 2,000 fs after laser irradiation starts in the case of a fluence of 74 $\mr{mJ/cm^2}$. 
   The electronic and atomic/lattice temperatures at the same depth start to differ immediately after the laser irradiation (160 fs);
   however they converge at 2,000 fs.
   At the boundary of the 3D TTM-MD and 1D TTM regions, the temperatures in the two regions are smoothly connected.}
   \label{fig:T-dist}
  \end{figure}

  %...Typical simulation result
  First, we describe a typical result of laser ablation obtained by TTM-MD.
  The laser energy injected from the surface is converted to the $\Te$ distribution along $x$ (depth direction), and it produces the temperature difference between $\Te$ and $T_\mr{a}$.
  The temperature difference causes energy transfer from the electronic system to the atomic system according to Eqs.~\ref{eq:3d-ttm} and \ref{eq:Langevin}.
  Figure \ref{fig:T-dist}(a) and (b) show the temperature profiles $\Te(x)$ and $T_\mr{a}(x)$ in the case of a fluence of 74 mJ/cm$^2$ immediately after the incident laser stopped and after the temperature difference vanished, respectively.
  $\Te$ increases close to or over 20,000 K near the surface owing to the high energy density of the strong short-pulse laser and quickly decreases until $\Te(x)$ equilibrates with increasing $T_\mr{a}(x)$ at any $x$.
  After the two temperatures equilibrate, the temperatures $\Te$ and $T_\mr{a}$ gradually diffuse to the deeper region treated by 1D-TTM.
  At $x=200$ nm in Fig.~\ref{fig:T-dist}, the two TTM systems, 3D-TTM and 1D-TTM, are smoothly connected and well simulating the thermal diffusion from the surface to the deep region.

  %...Non-reflecting boundary condition
  Figures \ref{fig:depth-time} show the colour-coded time evolution of the surface with according to the atomic temperature.
  The atomic system is truncated at approximately -190 nm.
  The shockwaves produced by the laser are shown by red bands toward the bottom, while those that reached the bottom did not reflect.
  This shows that the NRBC works well in absorbing the strong shockwaves.
  Thus, it is confirmed that this hybrid 3D-TTM-MD and 1D-TTM approach can efficiently perform large-scale laser-ablation MD simulation with a reduced number of atoms existing only close to the surface.

  \subsection{$\Te$-effect on TTM-MD laser ablation}\label{sec:Te-ttm}

\begin{figure*}[tbp]
   \centering
   \includegraphics[keepaspectratio,width=\textwidth]{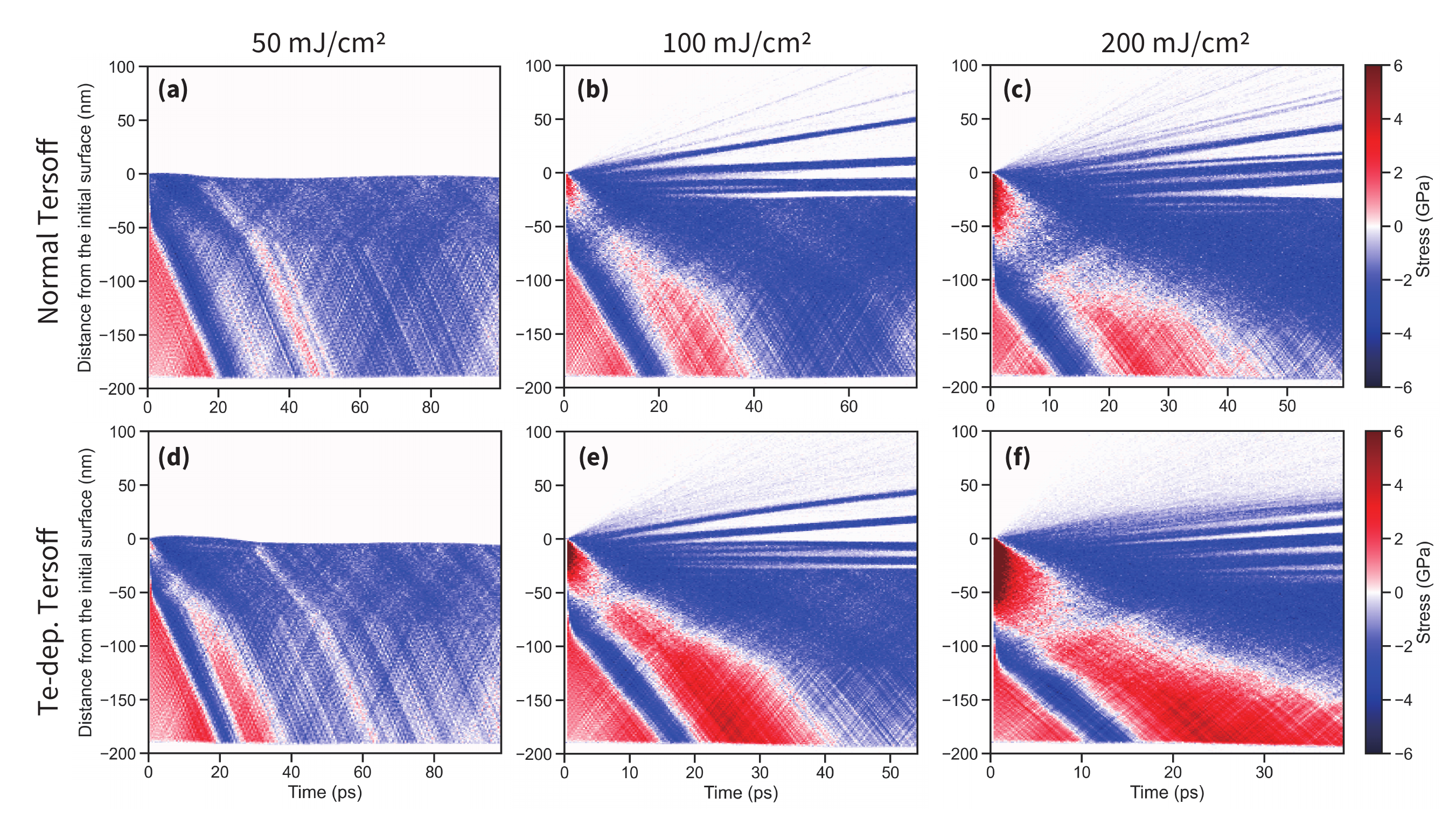}
   \caption{Comparison of the laser ablation MD simulation. Notice that the time ranges are different plot by plot.}
   \label{fig:depth-time}
  \end{figure*}

  %...Te-depによって何がどの程度変化するか
  Figures \ref{fig:depth-time}(a)-(f) compare the results of the laser ablation TTM-MD simulation of several laser fluences between the normal and $\Te$-dependent potentials.
  The most noticeable difference is the compressive pressure near the surface immediately after the laser irradiation.
  The difference in the maximum pressure near the surface after the laser irradiation at a fluence of 100 mJ/cm$^2$ is 12.7 GPa for the $\Te$-dependent potential and 3.2 GPa for the normal Tersoff potential; that is, an approximately four-times-greater pressure is caused by the $\Te$ effect on the potential.
  The other fluences have a similar tendency, and a 3-4-times-greater pressure appears in the case of the $\Te$-dependent potential.

  %...No effects on ablation
  Although the pressure caused by the laser irradiation is considerably different, subsequent processes are not significant.
  For example, comparing Figs.~\ref{fig:depth-time}(b) and \ref{fig:depth-time}(e) (fluence = 100 mJ/cm$^2$), in both cases, after approximately 10 ps, several spallation-like events occur immediately below the surface, and blocks of thickness 5-10 nm are flown away.
  The timing of spallation events and the thickness of blocks are comparable.
  This is counter-intuitive because the tensile pressure is considered to be a consequence of the reflection of the compressive pressure wave at the surface, and thus, the strength of the tensile pressure is expected to be proportional to the compressive pressure created by the laser irradiation.\cite{Paltauf1996-an}
  This can be explained by the fact that the equilibrium volume of melt, $V_\mr{eq}^\mr{melt}$, at a certain temperature is smaller than that of the crystalline diamond structure, $V_\mr{eq}^\mr{crystal}$, at the same temperature.
  As shown in Fig.~\ref{fig:Tersoff}(b), both values of $V_\mr{eq}$ increase with increasing $\Te$, and this $\Te$ dependence of $V_\mr{eq}$ contributes to the increase of the compressive pressure.
  However, because $V_\mr{eq}^\mr{melt}$ is smaller than $V_\mr{eq}^\mr{crystal}$ at the same $T_\mr{a}$ and $\Te$, the higher compressive pressure enhances melting and releases the pressure.
  This mechanism cannot occur in other materials with $V_\mr{eq}^\mr{melt} > V_\mr{eq}^\mr{crystal}$; hence, the $\Te$ effect on the tensile pressure (and thus the spallation-like events) can be remarkable for most metallic materials.

\begin{figure}[tbp]
   \centering
   \includegraphics[keepaspectratio,width=.5\textwidth]{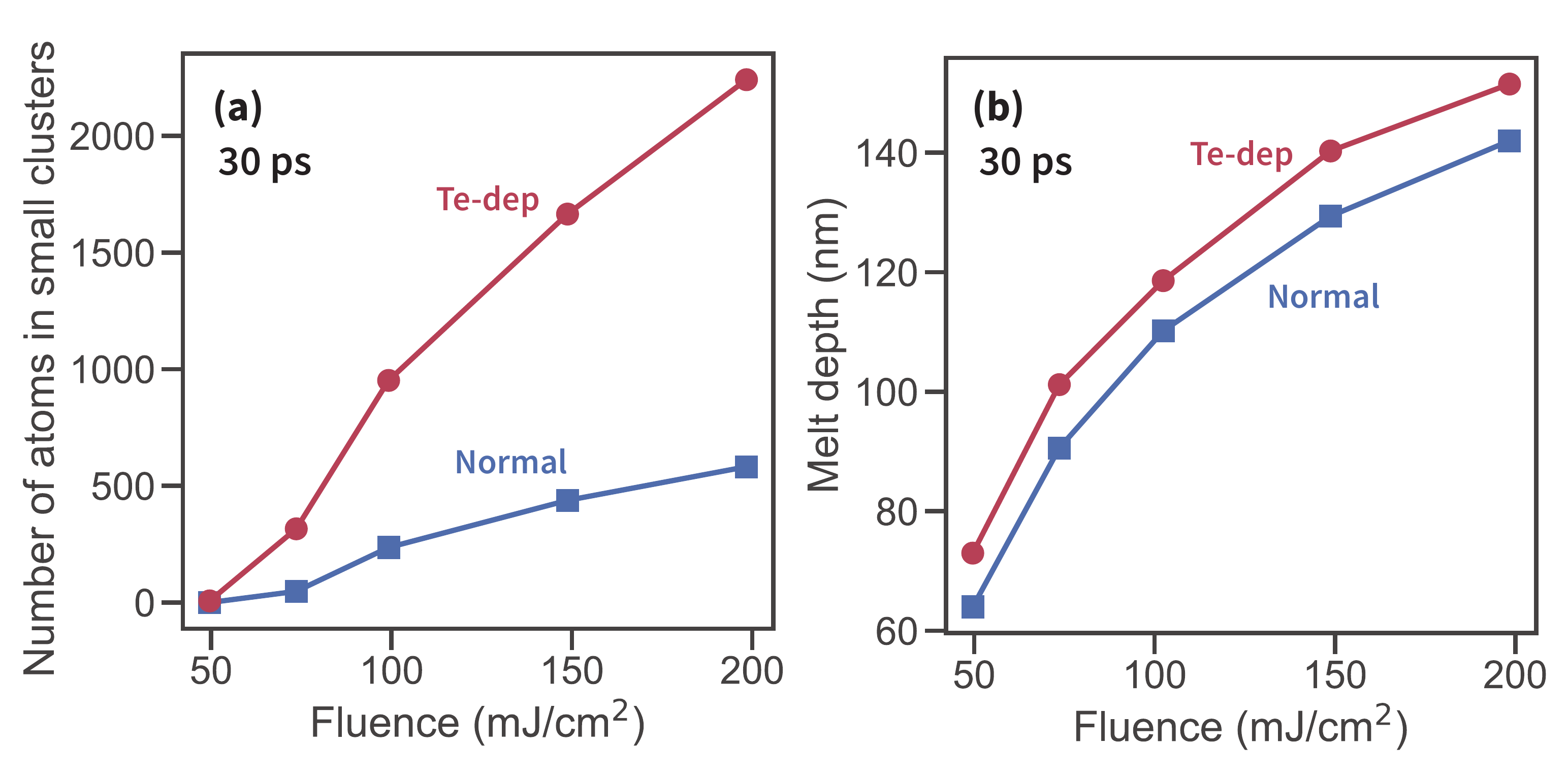}
   \caption{(a) Number of atoms flown away from the surface as clusters of less than 10 atoms and
   (b) melt depth from the initial surface as a function of the laser fluence.}
   \label{fig:vs-fluence}
  \end{figure}

  %...小さなクラスタとしての飛散量（4x程度）
  Compared to spallation, the $\Te$ effect on evaporation is more significant.
  A comparison of Figs.~\ref{fig:depth-time}(c) and \ref{fig:depth-time}(f) indicates that evaporated atoms assume cluster forms in the case of the normal potential.
  By contrast, they assume atomic or smaller-size cluster forms in the case of the $\Te$-dependent potential.
  Figure \ref{fig:vs-fluence}(a) shows the number of evaporated atoms in small clusters (including <10 atoms) determined from the snapshot at 30 ps.
  There are approximately four times more evaporated atoms within small clusters in the case of the $\Te$-dependent potential in all cases of different fluences.
  This is reasonable because $\Te$ becomes very high at the top of the surface, as shown in Fig.~\ref{fig:T-dist}(a), and the $\Te$-dependent potential changes the property of Si significantly for high $\Te$, as shown in Fig.~\ref{fig:Tersoff}.

  %...Evaporation of spalled blocks
  The promotion of atomic evaporation because of high $\Te$ causes a qualitatively different phenomenon that is not observed in the case of the normal potential.
  As the top-most spalled block has high $\Te$ in the case of high fluence, the evaporation of atoms continues even after the block is separated from the original surface and the evaporated atoms kick back the block downward, resulting in the slow-down of the spalled block.
  The block either gets together with a trailing block or evaporates completely in the end.

  %...melt depth (10nm程度)
  Figure \ref{fig:vs-fluence}(b) shows the melt depth from the initial surface at 30 ps.
  Clearly $\Te$ enhances melting, and thus, the melt depth increases by approximately 10 nm in the $\Te$-dependent case.
  Note that, however, we did not observe any significant difference in the density and local structure of the melt between the cases of the normal and $\Te$-dependent potentials.

 \section{Conclusion}\label{sec:conclusion}

 In this study, we performed the two-temperature model (TTM)-MD simulation of laser ablation to investigate the effects of the electronic temperature ($\Te$) on the ablation response of the Si surface.
 By comparing the MD results obtained with the normal and $\Te$-dependent Tersoff IPs, the $\Te$ effects on the laser ablation of Si can be summarised as follows.
 \begin{itemize}
  \item An approximately four-times-greater compressive pressure is generated below the surface a few ps after the laser irradiation because of the strong dependence of the equilibrium volume on $\Te$.
  \item Unlike the strong compressive pressure, the tensile pressure generated after the compressive shockwave leaves the surface is not different between the normal and $\Te$-dependent cases. Thus, the spallation phenomena caused by the tensile pressure are similar in both cases.
  \item A high $\Te$ at the very top of the surface enhances the evaporation of atoms, thereby increasing the number of evaporated atoms or small clusters. These evaporated atoms kick back the spalled block and thereby slows it down.
  \item The $\Te$-dependent potential slightly enhances melting, and the melt depth increases by approximately 10 nm in all cases of fluence. The structural properties of the melt do not seem to be affected by $\Te$.
 \end{itemize}
 These conclusions may not be applicable to other materials because the $\Te$ dependence of IPs can significantly differ owing to the difference in the electronic density of states from material to material.
 Particularly, the tensile pressure near the surface after the strong compressive wave leaves can be significantly different from the present case when the equilibrium volume of the melt is not smaller than that of the crystal structure as in the case of materials such as Si.
 The relationships between the electronic structure of the materials, the $\Te$ dependence of potential, and response to laser irradiation must be clarified in future studies.

 \section*{Acknowledgments}

 This work was supported by the MEXT Quantum Leap Flagship Program (MEXT Q-LEAP) under Grant No. JPMXS0118067246.
 The computations in this study were conducted using the supercomputer, Flow, at Information Technology Center in Nagoya University.
 The authors would like to thank Dr. Tanaka and Prof. Tsuneyuki at Tokyo University for the fruitful discussion,
 Prof. Ono at Gifu University for providing us with useful information about the $\Te$-dependent potential,
 and Editage (www.editage.com) for English language editing.

 \appendix
 \section{Algorithm for connecting 3D-1D TTM systems}

 The grid sizes in 3D and 1D TTM systems are different,
 and the boundary between these systems changes depending on the position of the right-most atoms
 that serve as the NRBC.
 Thus, this section explains a special treatment for connecting these two systems.
 
\begin{figure}[tbp]
   \centering
   \includegraphics[keepaspectratio,width=.49\textwidth]{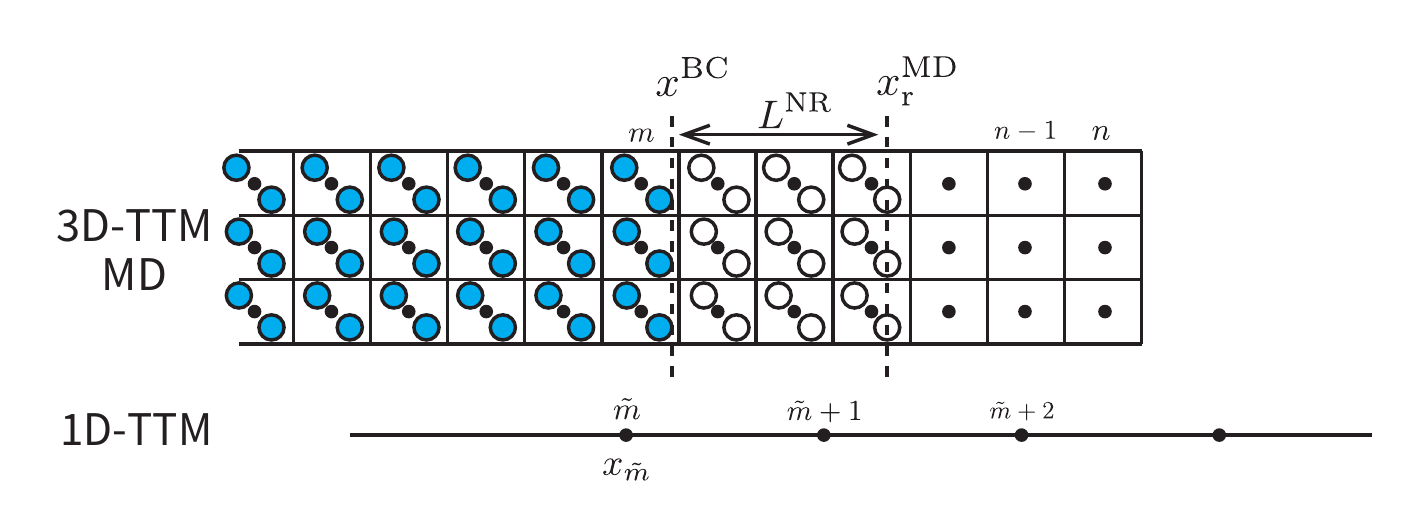}
   \caption{Configuration around the boundary of 1D and 3D TTM regions. 
   Atoms are drawn as circles. Open circles denote the atoms used for the NRBC.
   Black dots inside the lattice in the 3D-TTM system and on the line in the 1D-TTM system are grid points used to solve these TTM systems.}
   \label{fig:TTM-BC}
  \end{figure}

 Figure \ref{fig:TTM-BC} shows the configuration around the boundary of the 3D- and 1D-TTM systems.
 The tilde indicates that the variable with it is associated with the 1D-TTM system.
 In every MD step after atom positions are updated, the boundary $x$-position is determined by $x^\mr{BC} = x^\mr{MD}_\mr{r} -L^\mr{NR}$,
 where $x^\mr{MD}_\mr{r}$ and $L^\mr{NR}$ are the right-most atom position and length of the NRBC, respectively.
 The grid points in the 1D-TTM system to which the boundary condition is applied are $x_{\tilde{m}} < x^\mr{BC}$.
 The electronic and lattice temperatures at $x_{\tilde{m}}$ are given as $\tilde{T}_\mr{c}(x_{\tilde{m}}) = \langle T_\mr{c}(x_m)\rangle$,
 where $x_m$ is the grid point in the 3D-TTM system closest to $x_{\tilde{m}}$, $\langle \cdot \rangle$ denotes averaging over the $y$- and $z$-directions, and the subscript c stands for the carrier of the temperature e (electron) or l (lattice).

 The boundary conditions for the electronic and lattice temperatures in the 3D-TTM system are given as:
 \begin{equation}
  T_\mr{c}(x_i) = \tilde{T}_\mr{c}(x_{\tilde{k}}) +\frac{x_i -x_{\tilde{k}}}{x_{\tilde{k}+1} -x_{\tilde{k}}} \left[ \tilde{T}_\mr{c}(x_{\tilde{k}+1}) -\tilde{T}_\mr{c}(x_{\tilde{k}})\right],
 \end{equation}
 where $m<i \leq n$, $n$ is the number of mesh points along $x$ in the 3D-TTM system,
 and $x_{\tilde{k}} \leq x_i < x_{\tilde{k}+1}$.
 The temperature of the grid point $i$ in the 3D-TTM system is determined by the linear interpolation of the temperature in the 1D-TTM system at $x_i$.
 Similarly, the electron and lattice temperatures at $x>x^\mr{BC}$ in the 3D-TTM system are determined by the 1D-TTM system,
 and conversely, those at $x<x^\mr{BC}$ in the 1D-TTM system are determined by the 3D-TTM system,
 resulting in a smooth connection between the two systems.
 
 \section*{References}
 \bibliographystyle{iopart-num}
 \bibliography{reference}

\end{document}